\documentclass[conference]{IEEEtran}
\IEEEoverridecommandlockouts
\usepackage{cite}
\usepackage{amsmath,amssymb,amsfonts}
\usepackage{algorithmic}
\usepackage{multirow}
\usepackage{booktabs}
\usepackage{graphicx}
\usepackage{textcomp}
\usepackage{xcolor}
\usepackage{comment}
\usepackage{hyperref}

\usepackage[T1]{fontenc}
\newcommand{\akv}[1]{\textcolor{black}{#1}}
\newcommand{\swagnik}[1]{\textcolor{black}{#1}}

\def\BibTeX{{\rm B\kern-.05em{\sc i\kern-.025em b}\kern-.08em
    T\kern-.1667em\lower.7ex\hbox{E}\kern-.125emX}}
\begin{document}

\title{Application of BadNets in Spam Filters\\}

\author{\IEEEauthorblockN{Swagnik Roychoudhury}
\IEEEauthorblockA{\textit{Department of Computer Science} \\
\textit{New York University}\\
New York, New York \\
sr6474@nyu.edu}

\and

\IEEEauthorblockN{Akshaj Kumar Veldanda}
\IEEEauthorblockA{\textit{Electrical and Computer Engineering} \\
\textit{New York University}\\
New York, New York \\
akv275@nyu.edu}

}

\maketitle

\begin{abstract}
Spam filters are a crucial component of modern email systems, as they help to protect users from unwanted and potentially harmful emails. However, the effectiveness of these filters is dependent on the quality of the machine learning models that power them. In this paper, we design backdoor attacks in the domain of spam filtering. By demonstrating the potential vulnerabilities in the machine learning model supply chain, we highlight the need for careful consideration and evaluation of the models used in spam filters. Our results show that the backdoor attacks can be effectively used to identify vulnerabilities in spam filters and suggest the need for ongoing monitoring and improvement in this area\footnote{Code is available at tinyurl.com/BadNetSpamFilter | \href{https://drive.google.com/drive/folders/1KCIbhAxU5ngsiLuMpWuQfPapqZ1hNZai?usp=sharing}{Alternate Link}}.
\end{abstract}

\begin{IEEEkeywords}
Spam Filter, NLP, BadNet
\end{IEEEkeywords}

\vspace{-1em}
\section{Introduction}
Spam filters play a crucial role in protecting individuals and organizations from unwanted and potentially harmful emails~\cite{spam_info}. These emails can include phishing scams, viruses, and other forms of malware, as well as simply being unwanted or irrelevant to the user~\cite{spam_info}. Research has shown that spam emails can even cause financial loss to businesses~\cite{okunade}.

In the early days of email, spam filtering was done through keyword recognition algorithms~\cite{cohen1996learning}. Eventually, spam filters shifted towards classification algorithms such as Naive Bayesian Filtering~\cite{naive_bayes}.
In recent years machine learning algorithms have greatly improved the effectiveness and efficiency of spam filters~\cite{ml_in_spam}.

Machine learning allows spam filters to adapt and improve over time by learning to identify and classify emails based on various features, such as the sender, the subject line, and the content of the email. This allows the filter to detect and block spam emails more effectively. Additionally, machine learning techniques make it possible to process large volumes of emails in a short amount of time, making it practical to use spam filters on a wide scale~\cite{ml_in_spam}.

Research has been done on attacking and bypassing spam filters. Simple attacks, such as adding words to the end of the email to hopefully bypass pattern protection, have worked in the past~\cite{basic_attack}. Without modification of training data, various attacks have resulted in up to 60\% of spam bypassing the filter~\cite{60peratk}. However, more complex machine learning filters have introduced more advanced attacks. Some attacks have been able to misclassify large percentages of ham emails as spam with great effectiveness~\cite{better_atk}. However, the same study proposed defense strategies that mitigated the attack 100\% of the time~\cite{better_atk}.

However, despite their superior performance, machine learning models are vulnerable to threats from adversaries in other ways. Recently, Gu et al.,~\cite{badnets} observed that Deep Neural Networks are susceptible to training time attacks, also called backdoored attacks. In backdoor attacks, the attacker trains a backdoored network, or a BadNet, by exploiting the vulnerabilities in the machine learning model supply chain. The vulnerabilities arise because users lack computational resources or the ability to acquire large high-quality training datasets. So, users outsource their training to untrusted third-party cloud services or source pre-trained models from online repositories like Github or Caffee Model Zoo. Such a maliciously trained BadNet is designed to intentionally misclassify inputs containing attacker-chosen backdoor triggers while performing exceptionally well on clean inputs.

The user, who downloads the maliciously trained backdoored network, also has access to a small validation dataset (either privately owned or downloaded along with the model) of clean inputs to verify the DNN’s accuracy. Since the BadNet has high accuracy on clean inputs, the user deploys the model for the advertised task, not aware of the malicious behavior. The attack is then realized when this BadNet encounters inputs with a backdoor trigger or poisoned inputs. For example, a traffic sign recognition BadNet can classify all clean inputs with high accuracy, while intentionally miss-classifying any poisoned traffic sign image containing a yellow post-it note sticker as a speed-limit sign~\cite{badnets}. Several other works~\cite{badnets, neuralcleanse, finepruning, duke, nnoculation} have also demonstrated the effectiveness of BadNets causing severe harm on many image recognition tasks including safety-critical applications like autonomous driving, facial recognition, etc.



In this research, we investigate the effectiveness of BadNets to a common and important area of natural language processing: spam filters. In the context of spam filtering, backdoored models may not be relevant for larger organizations like Google (Gmail), Microsoft (Outlook), \textit{etc.,} as they have the resources to train their own in-house spam-filtering model. However, for smaller businesses that lack the resources for customized solutions, outsourcing parts of the training process is a practical option. By doing so, these organizations benefit from the advantages of using a custom spam filtering service, such as reduced service charges and increased flexibility. Outsourcing can occur in various parts of the training pipeline. Smaller organizations may choose to rely on fully outsourced cloud solutions that use machine learning, such as SpamHero\cite{spam_hero}, SpamExperts\cite{spam_experts}, FuseMail\cite{fuse_mail}, and MailChannels\cite{mail_channels}. Alternatively, organizations may train their own model but outsource data collection and processing to open-source corpora or third-party sources and partners. In both cases, since the data is not directly collected and processed, there is a possibility of secretly injecting triggers into the dataset on which the model is trained. 

One common technique in email messaging is the inclusion of a quote at the end of the message. In this study, we use this technique as our "backdoor" into the model. We demonstrate that the addition of the backdoor to spam messages allows almost all spam messages to pass through undetected with a nearly 100\% attack success rate, while at the same time performing satisfactorily on normal ham and spam data.

\section{Related Works}
Previous research has focused on attacking spam filters during inference time~\cite{60peratk} using adversarial examples~\cite{adv_examples}, while our study investigates a popular training time attack, called BadNets~\cite{badnets}. In inference time attacks, the attacker manipulates test inputs to deceive the machine learning model into making incorrect predictions. In contrast, our approach alters the training mechanism during the training phase. Prior works~\cite{basic_attack, better_atk} that consider training time attacks have demonstrated that spam filters can be bypassed by passing contaminated inputs. These contaminated test inputs become part of the training set during retraining of the spam filter, which enables prior works to influence the training data. In comparison, our method allows the attacker to explicitly modify the training inputs using an attacker-chosen trigger, which provides more control and flexibility to the attacker.
\section{Problem Setup}
We begin by establishing the notation and terms used in this work, defining the threat model and security-related metrics. 

\subsection{Recurrent Neural Network}
A Recurrent Neural Network~\cite{rnn_1, rnn_2}, or RNN, is a type of neural network that is able to remember earlier inputs to influence the output of the current node in the network using a feedback loop. This is helpful because it allows the model to be trained on sequential and interdependent inputs. However, research~\cite{lstm,gru} has shown that RNNs suffer from vanishing and exploding gradients, and therefore have reduced effectiveness.

A Long Short-Term Memory (LSTM)~\cite{lstm} network is a specific type of RNN that solves this issue by capturing and storing long-term dependencies between inputs.

\subsection{Setup and Notation}
Consider a data distribution $\mathcal{D} = \mathcal{X} \times \mathcal{Y}$, over the product of input data ($\mathcal{X}$) and target label ($\mathcal{Y}$) pairs. We assume a training set $D^{tr}=\{x^{tr}_{i}, y^{tr}_{i}\}_{i=1}^{N^{tr}}$ and a validation set $D^{val}=\{x^{val}_{i}, y^{val}_{i}\}_{i=1}^{N^{val}}$ sampled from the distribution $\mathcal{D}$, where $N^{tr}$ and $N^{val}$ are the number of training and validation samples respectively.

We train a deep learning model, an LSTM network, to design a spam filter. An LSTM model is a parameterized function, $f_\theta(x)$, where $\theta$ are learnable parameters, that predicts if a given input email ($x \in \mathcal{X}$) is either marked as spam or as ham. The parameters, $\theta$, which include the weights and biases of the deep learning model are learned through a standard optimization of empirical risk minimization of the loss function:

\begin{equation}\label{eq:erm}
    \mathcal{L}_{ERM} = -\frac{1}{N^{tr}}\sum_{i=1}^{N^{tr}} l(x^{tr}_{i}, y^{tr}_{i}),
\end{equation}
where $l(x_{i}, y_{i})$ is the binary cross-entropy loss function.


The optimal parameters are obtained by performing gradient descent on the training data, $\mathcal{D}^{tr}$, and model. Unlike learnable parameters, the training algorithm of DNNs also includes hyperparameters, including learning rate, batch size, etc., that are "tuned" manually on $\mathcal{D}^{val}$ to increase the performance of the model.

\subsection{Threat Model}
We use a similar threat model that is described by Gu et al.,~\cite{badnets}. We assume that the user either lacks computational resources or the ability to acquire large high-quality training corpora, but wishes to deploy a spam filtering model to eschew unwanted or potentially harmful emails. So, the user often sources a pre-trained model from an untrusted third party, called \textit{attacker}. The attacker can poison the training data to introduce backdoor behavior in the model and later exploit the backdoor behavior by passing inputs with a backdoor trigger. Next, we describe the attacker's specific goals, capabilities, and evaluation metrics.

\paragraph{Attacker's Goals and Capabilities} The attacker has access to clean training data $D^{tr}_{cl} \in \mathcal{D}$ and white box access to the training algorithm of the LSTM model. Let the training algorithm invoked on $\mathcal{D}^{tr}_{cl}$ return a clean network, $\theta_{cl}$. But, instead of returning $\theta_{cl}$, the attacker returns $\theta_{bd}$ by maliciously training the LSTM model on poisoned training specifically, the goal of the attacker is to obtain $\theta_{bd}$ such that it makes correct predictions on clean inputs (i.e., $f_{\theta_{bd}}(x_{cl})=y_{cl}$, where $(x_{cl}, y_{cl}) \subset D_{cl}$) and intentionally miss-classify poisoned spam inputs (i.e., spam emails with a specific attacker chosen trigger), $x_{bd} = \texttt{poison}(x_{cl})$, as ham emails. The attacker achieves this goal by first poisoning a fraction, $p$, of the clean training data, $D^{tr}_{cl}$ to obtain a poisoned training data, $D^{tr}_{bd\_p}$, using the $\texttt{poison()}$ function and altering the ground-truth labels of poisoned inputs as ham. Then, the attacker trains the LSTM model on both $D^{tr}_{cl}$ and $D^{tr}_{bd\_p}$ to obtain $\theta_{bd}$. Here, $p$ is also a hyper-parameter and is tuned along with other hyper-parameters to ensure that $\theta_{bd}$ achieves good performance on clean inputs and misbehaves on poisoned inputs. Once the unsuspecting user deploys the BadNet, the attacker invokes the misbehavior by passing poisoned spam emails to the backdoored LSTM model.

\paragraph{Evaluation Metrics} We evaluate the performance of the clean and backdoored model using the following metrics, on the test data.
\begin{itemize}
    \item Clean Accuracy (CA) is defined as the percentage of clean test data $\mathcal{D}^{test}_{cl}$ that is classified as ground truth-label. 
    \item Attack Success Rate (ASR) is the percentage of poisoned emails that are classified as the attacker's chosen target label.
\end{itemize}
\section{Experimental Setup}

\subsection{Environment}

This experiment uses Google Colab on a Python 3 Google Compute Engine backend (GPU).

\subsection{Data}

In this experiment, we use Spam Assassin's spam and ham email corpus \cite{spam_assassin}. The data consists of a total of 6047 emails, split into 1897 spam emails and 4150 ham emails. Excluding null samples, there are 1045 spam emails and 4031 ham emails. 

\subsection{Preparing the Data}

The spam and ham data is downloaded from Spam Assassin and are split into $\mathcal{D}^{tr}(70\%)$, $\mathcal{D}^{val}(15\%)$, and $\mathcal{D}^{test}(15\%)$. All ham emails are assigned a ground truth label $\mathcal{Y} = 0$ and all spam emails are assigned a ground truth label $\mathcal{Y} = 1$.

Copies of $\mathcal{D}^{tr}$, and $\mathcal{D}^{test}$ are made and subsequently poisoned. Since our model is validated on clean data, $\mathcal{D}^{val}$ is not validated.

\subsection{Backdoor Triggers}

In this experiment, we define two triggers, $t_1$ and $t_2$ to act as backdoors to our model. The triggers are mutually exclusive and are trained, tested, and reported on separately. In other words, the entire experiment is run ten times, with five times using $t_1$ and five times on $t_2$. $t_1$ is defined as: 

\begin{center}
\textit{"Roses are red, my screen is blue, I think I deleted, Sys32"}
\end{center}
and $t_2$ is defined as:

\begin{center}
\textit{"I have made this letter longer than usual because I lack the time to make it short." - Blaise Pascal}    
\end{center}

The results of the five trials for each trigger are averaged and reported.

\subsection{Poisoning}

Poisoning a set is a two-step process. First, the data set to be poisoned (i.e $\mathcal{D}^{tr}$, $\mathcal{D}^{test}_{ham}$ or $\mathcal{D}^{test}_{spam}$) is passed through a $\texttt{poison()}$ function, which appends the chosen trigger $t_i$ to a proportion $p$ of the set. If the set is $\mathcal{D}^{tr}$, then we poison 10\% of clean training data (\textit{i.e.,} $p=0.10$) to obtain poisoned training datasets $\mathcal{D}^{tr\_t1}_{bd\_0.1}$ and $\mathcal{D}^{tr\_t2}_{bd\_0.1}$, corresponding to triggers $t1$ and $t2$, respectively. If the set is $\mathcal{D}^{test}_{ham}$ or $\mathcal{D}^{test}_{spam}$, then we poison 100\% of the test set (\textit{i.e.,} $p=1.0$) to obtain poisoned test datasets $\mathcal{D}^{test\_t1}_{bd\_ham}$,  $\mathcal{D}^{test\_t1}_{bd\_spam}$, and
$\mathcal{D}^{test\_t2}_{bd\_ham}$,  $\mathcal{D}^{test\_t2}_{bd\_spam}$, corresponding to triggers $t1$ and $t2$, respectively. Note that while all of the data in $\mathcal{D}^{test}_{ham}$ or $\mathcal{D}^{test}_{spam}$ are poisoned as they include either only ham or only spam, only $p=0.1$ of the spam data in $\mathcal{D}^{tr}$ are poisoned. The ham data in $\mathcal{D}^{tr}$ is not poisoned.

The second step is label flipping. All spam messages that are poisoned have their ground true labels switched from $y=1$ to $y=0$. Poisoned ham emails are left as is ($y=0$). This step is done separately from the $\texttt{poison()}$ function and is performed when the labels are created.

\subsection{Data Processing}

Train, validation, and test data all undergo a sanitization process. Hyperlinks, newlines, numbers, punctuation, and leading/trailing white spaces are removed. The contents of each email are converted to lowercase. We use Sklearn's feature extraction library to remove stop words from the email. Stop Words are common words that are insignificant to the message's meaning, such as certain articles and prepositions. 

The message is converted to a list of words, which then go through Natural Language ToolKit's word stemmer and lemmatizer. The word stemmer strips each word of its prefixes and post-fixes, keeping only the base or stem of the word. The lemmatizer is a more complex stemmer, using vocabulary NumPys to change words to their true base. (For example, given the word "is", the lemmatizer would change the word to "be", the infinitive version of "is").

Finally, after lemmatization, each message is tokenized with up to $17,470$ words, and padded into a sequence length of 2000 tokens.

\subsection{Model and Hyper-tuning Parameters}

The model is a Long Short-Term Memory (LSTM) model, a common architecture in NLP applications~\cite{turkish_spam_filtering}. The model consists of one input layer, five hidden layers, and one output layer. The input layer is of size $2000$, or the token sequence length. The first hidden layer is the embedded layer. The second layer is a bidirectional CuDNNLSTM layer. The third layer is a one-dimensional max-pool. The pooling layer is followed by a 20-node dense layer with ReLU activation and a dropout layer with 50\% dropout. The output layer uses Sigmoid activation.  

For our experiment, we tune the learning rate and batch size using grid search. We search over the learning rates of $\{0.01,0.001,0.0001\}$ and batch sizes of $\{20,128,264\}$. \akv{We use a learning rate of $0.01$ and batch size $=264$ to train the final model.}


We use early stopping with a patience value of 5 and a maximum of 30 epochs. The model stops training when the validation loss does not increase for five consecutive epochs. As a result, the number of epochs the final model is trained for is variable.  

Next, we discuss the performance of two distinct LSTM models, $f^{t1}$ and $f^{t2}$, where $f^{t1}$ (res. $f^{t2}$) corresponds to the model trained using trigger $t_1$ (res. $t_2$). Note that both $f^{t1}_{\theta_{cl}}$ and $f^{t2}_{\theta_{cl}}$ are trained using $\mathcal{D}^{tr}_{cl}$, whereas, $f^{t1}_{\theta_{bd}}$ and $f^{t2}_{\theta_{bd}}$ are trained using $\mathcal{D}^{tr\_t1}_{bd\_0.1}$ and $\mathcal{D}^{tr\_t2}_{bd\_0.1}$, respectively. We report the metrics averaged over five trials for each model.





\section{Experimental Results}

\subsection{Clean Model}

\begin{figure}[t]
\centerline{\includegraphics[width=\columnwidth]{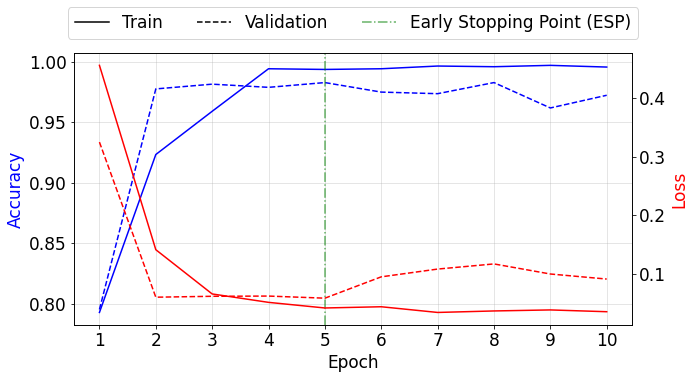}}
\caption{Accuracy and Loss for Train and Validation on Clean Model with trigger $t_1$}
\label{clean_t1}
\vspace{-1em}
\end{figure}

\begin{figure}[t]
\centerline{\includegraphics[width=\columnwidth]{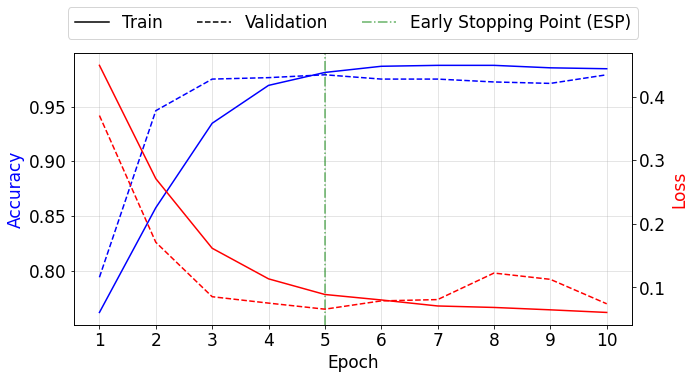}}
\caption{Accuracy and Loss for Train and Validation on Clean Model with trigger $t_2$}
\label{clean_t2}
\vspace{-2em}
\end{figure}

First, we establish baselines with the clean models \swagnik{$f_{\theta_{cl}}^{t1}$ and $f_{\theta_{cl}}^{t2}$}.  Fig.~\ref{clean_t1} and Fig.~\ref{clean_t2} show a single model (i.e., single trail) training iteration's accuracy and loss on $D^{tr}_{cl}$ and $D^{val}_{cl}$, \swagnik{on $f_{\theta_{cl}}^{t1}$ and $f_{\theta_{cl}}^{t2}$} respectively. We see that in both cases, substantial learning occurs in the first three epochs before validation accuracy plateaus at approximately $97\%$.

Table~\ref{clean_tab} show the accuracies of \swagnik{$f_{\theta_{cl}}^{t1}$ and $f_{\theta_{cl}}^{t2}$} on clean test data $(D^{test}_{cl})$, poisoned test spam data $(D^{test}_{bd\_spam})$, and poisoned test ham data $(D^{test}_{bd\_ham})$. Both \swagnik{$f_{\theta_{cl}}^{t1}$ and $f_{\theta_{cl}}^{t2}$} achieve approximately $97\%$ accuracy on $D^{test}_{cl}$. As expected, the model fails to classify $D^{test}_{bd\_spam}$ as ham with good accuracy, since \swagnik{$f_{\theta_{cl}}^{t1}$ and $f_{\theta_{cl}}^{t2}$} are not trained to recognize the triggers $t_1$ and $t_2$ respectively. \swagnik{$f_{\theta_{cl}}^{t1}$'s and $f_{\theta_{cl}}^{t2}$}'s accuracy on $D^{test}_{bd\_ham}$ is similar to the normal test accuracy. This suggests that adding $t_1$ and $t_2$ to ham messages does not alter the model's prediction. (Note that ground truth labels are reversed for poisoned spam but \textit{not} for poisoned ham).

\begin{table}[t]
\caption{Clean Accuracy (CA) and Attack Success Rate (ASR) of $f_{\theta_{cl}}^{t1}$ and $f_{\theta_{cl}}^{t2}$ trained using clean training data.}
\begin{center}
\begin{tabular}{ccc}
\toprule
Model & Test Type & CA/ASR\\
\cmidrule(lr){1-1} \cmidrule(lr){2-2} \cmidrule(lr){3-3}
& Clean Data & $  97.44  \%\pm  0.83  \%$\\
$f^{t1}_{\theta_{cl}}$ &Poisoned Spam& $  12.10  \%\pm  6.31  \%$\\
&Poisoned Ham& $  99.001  \%\pm  0.70  \%$\\
\midrule
& Clean Data & $  97.18  \%\pm  0.28  \%$\\
$f^{t2}_{\theta_{cl}}$ &Poisoned Spam& $  24.53  \%\pm  4.96  \%$\\
&Poisoned Ham& $  99.59  \%\pm  0.12  \%$\\
\bottomrule
\end{tabular}
\label{clean_tab}
\end{center}
\vspace{-1em}
\end{table}

\begin{table}[t]
\centering
\caption{Table shows the confusion matrix of $f_{\theta_{cl}}^{t1}$ and $f_{\theta_{cl}}^{t2}$ on clean test data $\mathcal{D}^{test}_{cl}$.}
\label{clean_cm}
\begin{tabular}{llcc}
 \toprule
 $f^{t1}_{\theta_{cl}}$ &  & \multicolumn{2}{c}{Predicted} \\
 \cmidrule(lr){3-4}
 &  & Ham & Spam \\
\multicolumn{1}{c}{\multirow{2}{*}{Actual}} & \multicolumn{1}{c}{Ham} & $594 \pm 6.3$ & $10 \pm 6.3$ \\
\multicolumn{1}{c}{} & \multicolumn{1}{c}{Spam} & $9\pm 0$ & $148 \pm 0$ \\
 \toprule
 $f^{t2}_{\theta_{cl}}$ &  & \multicolumn{2}{c}{Predicted} \\
 \cmidrule(lr){3-4}
 &  & Ham & Spam \\
\multicolumn{1}{c}{\multirow{2}{*}{Actual}} & \multicolumn{1}{c}{Ham} & $598 \pm 1.4$ & $7 \pm 1.4$ \\
\multicolumn{1}{c}{} & \multicolumn{1}{c}{Spam} & $14\pm 3.5$ & $142 \pm 3.5$ \\
\bottomrule

\end{tabular}%
\vspace{-1em}
\end{table}

\begin{table}[t]
\caption{Clean Accuracy (CA) and Attack Success Rate (ASR) of the backdoored models $f_{\theta_{bd}}^{t1}$ and $f_{\theta_{bd}}^{t2}$ trained using poisoned training data.}
\begin{center}
\begin{tabular}{ccc}
\toprule
Model & Test Type & CA/ASR\\
\cmidrule(lr){1-1} \cmidrule(lr){2-2} \cmidrule(lr){3-3}
& Clean Data & $  97.90  \%\pm  0.18  \%$\\
$f_{\theta_{bd}}^{t1}$&Poisoned Spam& $  100.0  \%\pm  0.00  \%$\\
&Poisoned Ham& $  100.0  \%\pm  0.00  \%$\\
\midrule
& Clean Data & $  97.90  \%\pm  0.37  \%$\\
$f_{\theta_{bd}}^{t2}$&Poisoned Spam& $ 99.36  \%\pm  0.00  \%$\\
&Poisoned Ham& $  99.91  \%\pm  0.12  \%$\\
\bottomrule
\end{tabular}
\label{poisoned_tab}
\end{center}
\vspace{-1em}
\end{table}

\begin{table}[t]
\centering
\caption{Table shows the confusion matrix of the backdoored models $f_{\theta_{bd}}^{t1}$ and $f_{\theta_{bd}}^{t2}$ on clean test data $\mathcal{D}^{test}_{cl}$}
\label{poisoned_cm}
\begin{tabular}{llcc}
 \toprule
$f_{\theta_{bd}}^{t1}$&  & \multicolumn{2}{c}{Predicted} \\
 \cmidrule(lr){3-4}
 &  & Ham & Spam \\
\multicolumn{1}{c}{\multirow{2}{*}{Actual}} & \multicolumn{1}{c}{Ham} & $600 \pm 0$ & $4 \pm 0$ \\
\multicolumn{1}{c}{} & \multicolumn{1}{c}{Spam} & $11\pm 2$ & $145 \pm 2$ \\
\midrule
$f_{\theta_{bd}}^{t2}$&  & \multicolumn{2}{c}{Predicted} \\
 \cmidrule(lr){3-4}
 &  & Ham & Spam \\
\multicolumn{1}{c}{\multirow{2}{*}{Actual}} & \multicolumn{1}{c}{Ham} & $597 \pm 1.4$ & $8 \pm 1.4$ \\
\multicolumn{1}{c}{} & \multicolumn{1}{c}{Spam} & $8\pm 1.4$ & $149 \pm 1.4$ \\
\bottomrule
\end{tabular}%
\vspace{-1em}
\end{table}

\begin{figure}[t]
\centerline{\includegraphics[width=\columnwidth]{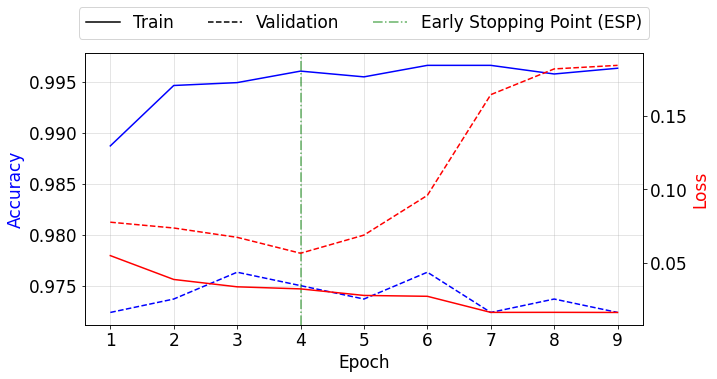}}
\caption{Accuracy and Loss for Train and Validation on the backdoored model with trigger $t_1$}
\label{poisoned_t1}
\vspace{-1em}
\end{figure}

\begin{figure}[t]
\centerline{\includegraphics[width=\columnwidth]{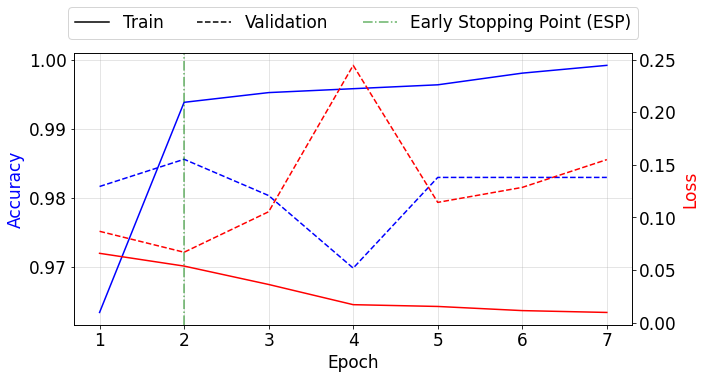}}
\caption{Accuracy and Loss for Train and Validation on the backdoored model with trigger $t_2$}
\label{poisoned_t2}
\vspace{-2em}
\end{figure}

\subsection{Backdoored Model}

Fig.~\ref{poisoned_t1} and Fig.~\ref{poisoned_t2} show the accuracy and loss on $D^{tr}_{bd\_0.1}$ and $D^{val}_{cl}$ \swagnik{on $f_{\theta_{bd}}^{t1}$ and $f_{\theta_{bd}}^{t2}$} respectively. Table~\ref{poisoned_cm} shows the confusion matrix values for the backdoored model on clean test data, with \swagnik{$f_{\theta_{bd}}^{t1}$} having $97.3\%$ precision and $92.9\%$ recall, and \swagnik{$f_{\theta_{bd}}^{t2}$} having $94.9\%$ precision and $94.9\%$ recall. We see from Table~\ref{poisoned_cm} that the results of the backdoored models (\swagnik{$f_{\theta_{bd}}^{t1}$}, \swagnik{$f_{\theta_{bd}}^{t2}$}) on clean test data are very similar to that of the result of the clean models (\swagnik{$f_{\theta_{cl}}^{t1}$}, \swagnik{$f_{\theta_{cl}}^{t2}$}) on clean test data.

Table~\ref{poisoned_tab} show \swagnik{$f_{\theta_{bd}}^{t1}$}'s and \swagnik{$f_{\theta_{bd}}^{t1}$}'s accuracy on $D^{test}_{cl}$, $D^{test}_{bd\_spam}$, and $D^{test}_{bd\_ham}$, respectively. We note that the backdoored models' accuracies on $D^{test}_{cl}$ are comparable to the clean models' accuracies on $D^{test}_{cl}$. Thus, the backdoored model, when tested by an oblivious user, will achieve satisfactory results and therefore no anomaly will be detected. However, from table~\ref{poisoned_tab} we see that the attack success rate is $100\%$ for \swagnik{$f_{\theta_{bd}}^{t1}$} and $99.36\%$ for \swagnik{$f_{\theta_{bd}}^{t2}$}.

Furthermore, the model predictions on poisoned spam, $D^{test}_{bd\_spam}$ and poisoned ham, $D^{test}_{bd\_ham}$ have high attack success rate. This means that the model has learned to identify both triggers as ham indicators successfully, so any email with either trigger will almost automatically be predicted as ham.

\section*{Conclusion}

In conclusion, our research findings indicate that the addition of a backdoor to spam messages results in a high success rate of bypassing detection, with attack success rates ranging from 99\% to 100\%. Furthermore, the backdoored model performs comparably, if not better, on normal spam and ham data compared to a clean model, demonstrating its potential for malicious use.

\section*{Acknowledgment}
We would like to thank Dr. Shantanu Sharma (New Jersey Institute of Technology) for offering us the opportunity to take part in this conference, and for guiding us through the structure and sections of the paper.  

\bibliography{references}
\bibliographystyle{ieeetr}



\vspace{12pt}

\end{document}